\documentclass[preprint]{emulateapj}       
\usepackage{amsmath,amssymb}
\usepackage{epsfig}
\usepackage{color}
\newcommand{\hMsol}{{\>h^{-1}\rm M}_\odot}
\newcommand{\hMpc}{{\>h^{-1}\rm  Mpc}} 
\newcommand{\hkpc}{{\>h^{-1}\rm kpc}}
\newcommand{\erfc}{\mathrm{erfc}}

\begin{document}
\title{The halo mass function conditioned on density from 
the Millennium Simulation: insights into missing baryons and galaxy mass functions}
%
\author {A.\ Faltenbacher\altaffilmark{1,2,3},     
  A.\ Finoguenov\altaffilmark{4,5} \& N.\ Drory\altaffilmark{4}} 

\altaffiltext{1}{ Max Planck Institut f\"ur  Astrophysik,   
  Karl-Schwarzschild-Str.   1,  85741  Garching, Germany } 
\altaffiltext{2}{  MPA/SHAO Joint Center for Astrophysical Cosmology   
  at  Shanghai   Astronomical  Observatory,   
  Nandan  Road 80,Shanghai  200030, China  } 
\altaffiltext{3}{  Physics Department, 
  University  of the  Western Cape,  
  Cape  Town 7535,  South Africa  }
\altaffiltext{4}{  Max-Planck-Institut f\"ur  Extraterrestrische Physik,
  Giessenbachstra\ss  e, 85748  Garching, Germany  } 
\altaffiltext{5}{ University  of  Maryland,  
  Baltimore  County, 1000  Hilltop  Circle,
  Baltimore, MD 21250, USA }
\begin{abstract}
  The  baryon  content of  high-density  regions  in  the universe  is
  relevant  to  two critical  unanswered  questions:  the workings  of
  nurture  effects on  galaxies  and the  whereabouts  of the  missing
  baryons. In this  paper, we analyze the distribution  of dark matter
  and  semianalytical   galaxies  in  the   Millennium  Simulation  to
  investigate  these  problems.    Applying  the  same  density  field
  reconstruction schemes  as used for the  overall matter distribution
  to  the matter locked  in halos  we study  the mass  contribution of
  halos  to  the  total   mass  budget  at  various  background  field
  densities,  i.e.,  the  conditional  halo  mass  function.  In  this
  context, we present a simple fitting formula for the cumulative mass
  function accurate to $\lesssim5\%$ for halo masses between $10^{10}$
  and  $10^{15}\hMsol$. We find  that in  dense environments  the halo
  mass function  becomes top  heavy and present  corresponding fitting
  formulae  for different  redshifts.  We  demonstrate that  the major
  fraction of matter in  high-density fields is associated with galaxy
  groups.  Since current X-ray surveys  are able to nearly recover the
  universal baryon  fraction within groups, our  results indicate that
  the  major part  of  the so-far  undetected warm--hot  intergalactic
  medium resides in low-density  regions.  Similarly, we show that the
  differences  in  galaxy  mass  functions with  environment  seen  in
  observed and  simulated data stem predominantly  from differences in
  the  mass distribution  of halos.   In particular,  the hump  in the
  galaxy mass function is  associated with the central group galaxies,
  and the bimodality observed in the galaxy mass function is therefore
  interpreted as that of central galaxies versus satellites.
\end{abstract}
\section{Introduction}
The  environmental  dependence  of  galaxy properties, such as broadband
color, star formation, and stellar mass, is a well-known effect in the
local universe \citep{Dressler-80}.  In this context environment means
an estimate  of the  smoothed density filed  at a given  location.  Of
particular interest  for the  present study is  the dependence  of the
galaxy   stellar   mass  function   (GMF)   on  environment.    Recent
comprehensive galaxy redshift surveys have led to intensive studies in
this field.   For instance, \cite{Mo-04}  model the dependence  of the
luminosity  function  on the  large-scale  environment  based on  mock
catalogs  for   the  the  two-degree  Field   Galaxy  Redshift  Survey
\citep[2dFGRS;][]{Colless-01} and \cite{Baldry-06} investigate  the GMF
as a function of environment in the nearby universe based on the Sloan
Digital Sky Survey  \citep[SDSS;][]{York-00}.  To uncover evolutionary
effects, surveys spanning a larger  redshift range have been used: the
study by \cite{Bundy-06} is based  on the DEEP2 Galaxy Redshift Survey
($0.4\leq  z \leq  1.4$);  \cite{Pannella-09} use  the COSMOS  survey;
\cite{Bolzonella-09} employ  the zCOSMOS survey in  the redshift range
$0<z<1$;   and   \cite{Scodeggio-09}   investigate   the   environment
dependence of the GMF based on the VVDS survey covering a redshift
range of $0.2 < z < 1.4$.

On intermediate  scales, $\sim\hMpc$, groups and  clusters of galaxies
themselves  provide a  definition  of environment.   At low  redshift,
\cite{Balogh-01}  using Two  Micron  All Sky  Survey  (2MASS) and  Las
Campanas Redshift Survey  (LCRS) data, separate different environments
as  field, groups, and  clusters, finding  that galaxy  luminosity and
mass functions depend on both, galaxy type (with steeper functions for
emission line galaxies)  and mass of the group  (with more massive and
brighter objects more common in  clusters), mainly as a consequence of
the   different   contribution   of   passive   galaxies   (see   also
\citealt{Hansen-09}).

The environmental  dependence of  properties of gas  is linked  to the
studies  of  warm--hot  intergalactic  media  (WHIM).   Hydrodynamical
simulations   by   \cite{Cen-Ostriker-99}   show  that   the   average
temperature of baryons is an increasing function of time, with most of
the baryons at  the present time having a temperature  in the range of
$10^5-10^7\rm  K$.   The  detection  of  this warm-hot  gas  poses  an
observational  challenge. While  according to  their census  more than
one-half of the  normal matter is yet to be detected,  it is not clear
which    methods   have   to    be   used.     In   a    later   study
\cite{Cen-Ostriker-06}  report that  the gas  density of  the warm-hot
intergalactic medium  is broadly  peaked at a  density three  to seven
times the critical density,  however their dark matter mass resolution
was only  moderate and a question  of assigning baryons  to groups has
not been addressed.

All the above  studies would greatly benefit from  a quantification of
the contribution of galaxy groups  to the density fields as a function
of  background density.  The  significance of  this question  has been
recognized  before  \citep[e.g.,][]{Sheth-Tormen-99},  but a  detailed
answer  has   never  been  provided.   Simulation   work  has  instead
concentrated  on  questions  regarding  halo  assembly  bias  and,  in
particular, on changes in the  properties of galaxies within the halos
of   similar    mass   but   residing    in   different   environments
\citep{Lemson-Kauffmann-99,                      Gao-Springel-White-05,
  Croton-Gao-White-07}.

This  paper is  organized  as follows.   In Section~\ref{sec:data}  we
review  the  Millennium  Simulation,  the  semi-analytic  modeling  of
galaxies   and   the   determination   of  the   background   density.
Section~\ref{sec:massfrac} examines the  dependence of the dark matter
halo  mass fractions  on  environment.  In  Section~\ref{sec:massfunc}
halo   mass  functions  in   different  environments   are  discussed.
Section~\ref{s:stars}  focuses  on  the  dependence  of  the  GMFs  on
environment.       We     present      a     short      summary     in
Section~\ref{sec:conclusion}.
\begin{figure*}
  \epsfig{file=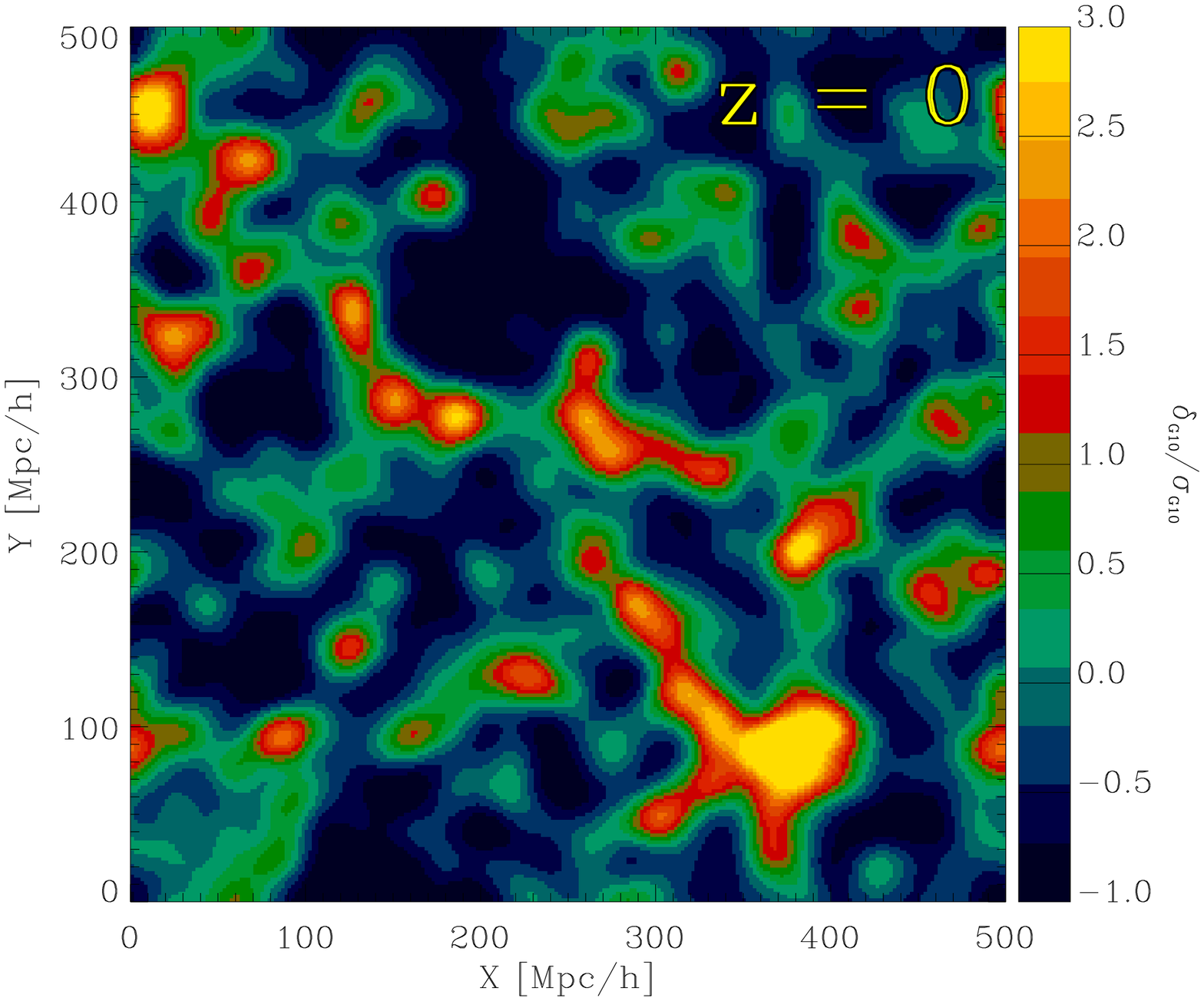,width=0.495\hsize}
  \epsfig{file=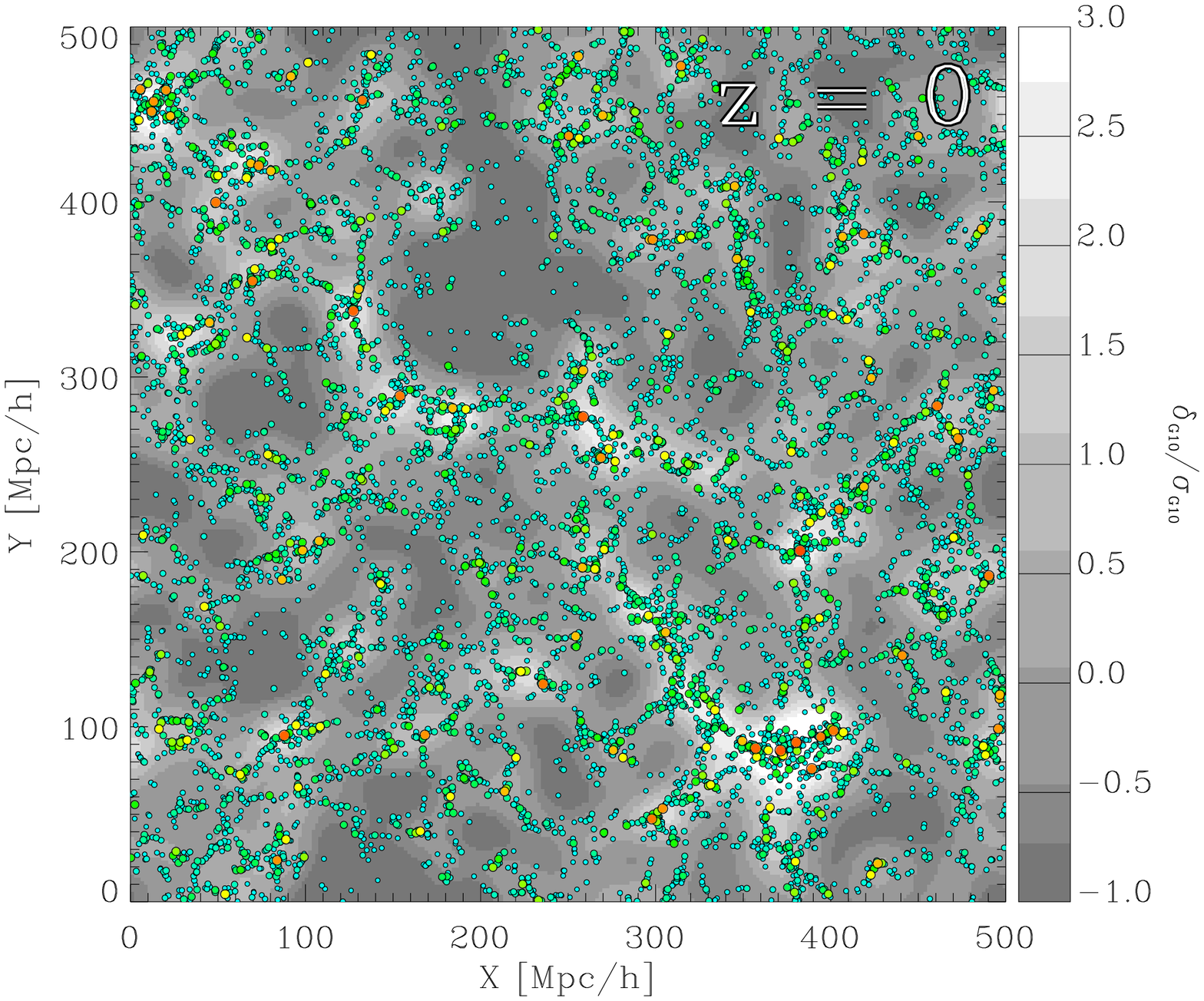,width=0.495\hsize}\\ 
  \epsfig{file=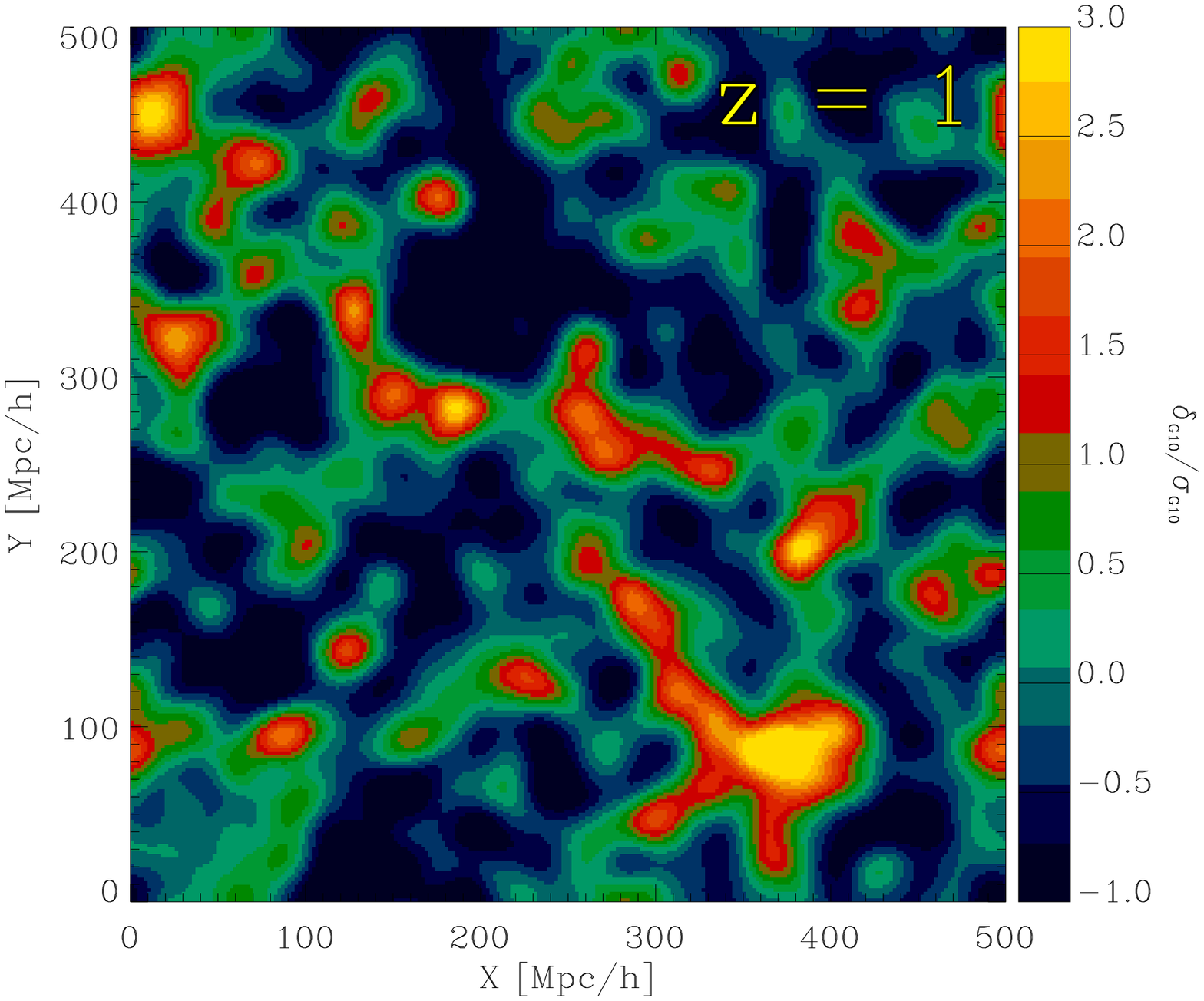,width=0.495\hsize}     
  \epsfig{file=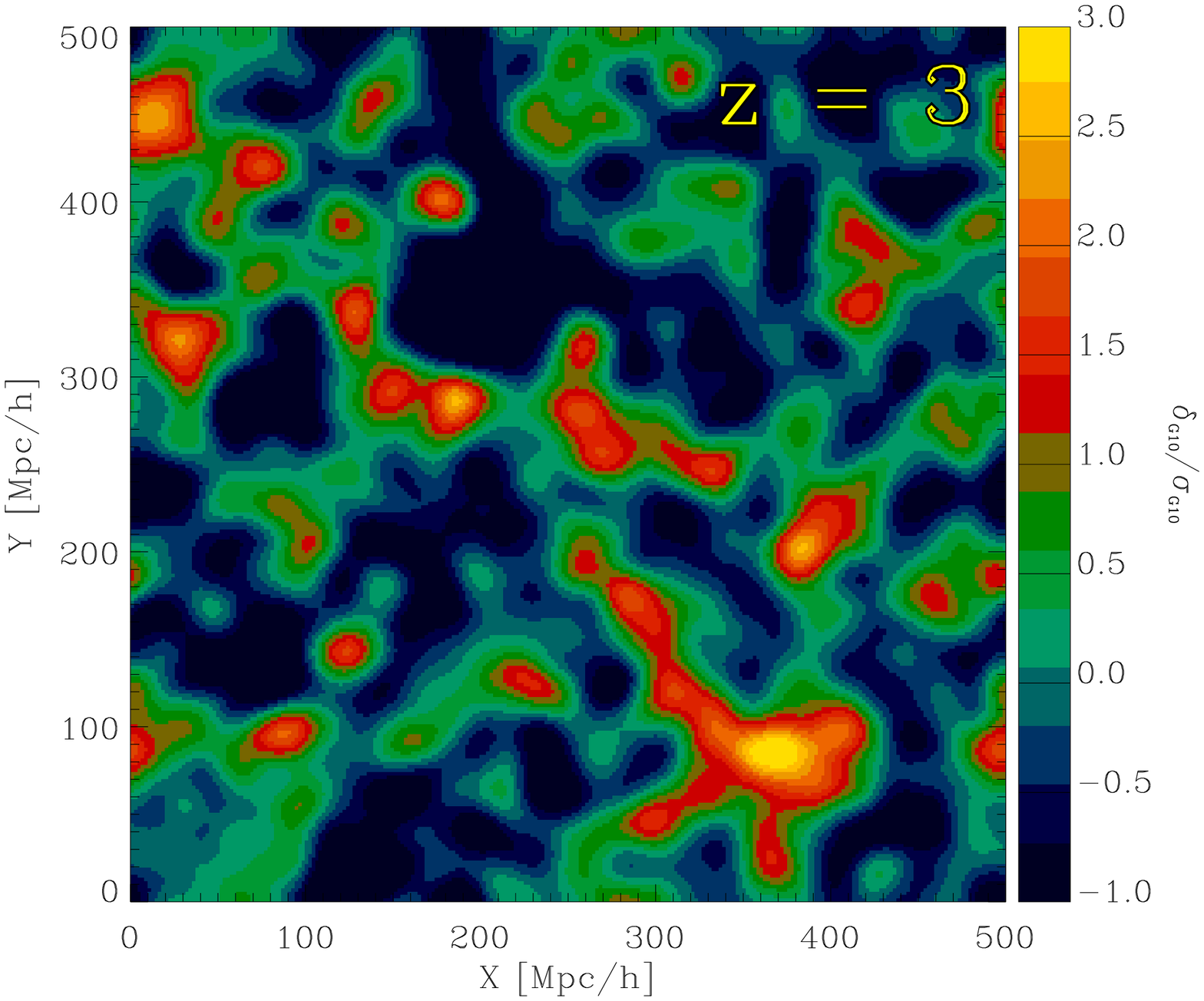,width = 0.495\hsize}
  \caption{\label{fig:step}  Upper  left  panel: dark  matter  density
    field at $z=0$, smoothed using  a Gaussian filter with a smoothing
    scale of $10\hMpc$, $\delta_{\rm  G10}$, and scaled by the average
    rms  fluctuations within  the  corresponding volume,  $\sigma_{\rm
      G10}$. Upper right panel: density  field as shown on the left in
    gray  scale.   The points  represent  halos above  $10^{12}\hMsol$
    within a slice of $10\hMpc$ thickness.  Colors (from green to red)
    and sizes (from small to large) correspond to the logarithmic mass
    of the halos.  As  expected, the halo distribution closely follows
    the  density  field. Lower  panels:  smoothed  and scaled  density
    fields,as the  upper left panel,  but for the redshifts  $z=1$ and
    $z=3$. According to linear  theory $\delta/\sigma$ does not change
    with time.}
\end{figure*}
\begin{figure}
  \epsfig{file=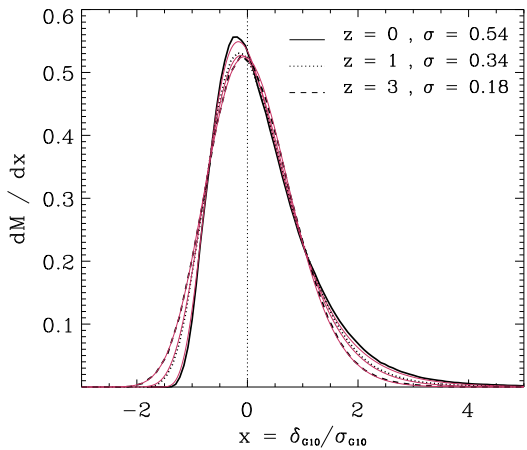,width = 0.95\hsize}
  \caption{\label{fig:dd} Differential mass distribution as a function
    of scaled  density contrast.  The  density is computed based  on a
    $10\hMpc$  Gaussian  smoothing  kernel,  and the  scaling  factor,
    $\sigma_{\rm G10}$,  is the  mass variance within  a corresponding
    volume.  Solid,  dotted, and  dashed lines indicate  the redshifts
    $z=0$,  $1$  and  $3$,  respectively.   The red  curves  show  the
    lognormal fits to the distribution.}
\end{figure}
\begin{figure}
\epsfig{file=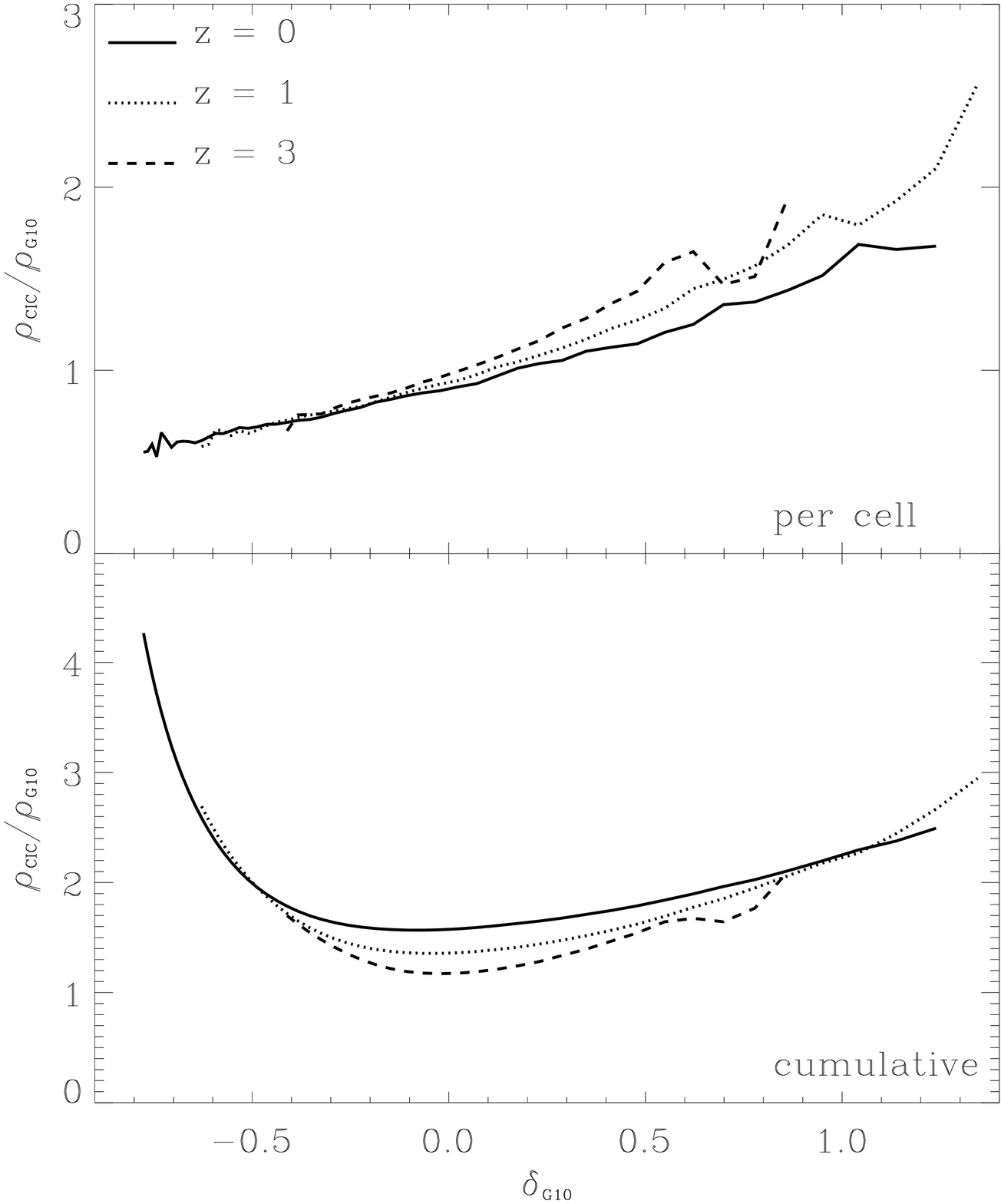, width = 0.95\hsize}
\caption{\label{fig:bg} Upper panel: mean CiC density as a function of
  smoothed  density.  Lower  panel:  mean density  within  isodensity
  contours divided by that isodensity.}
\end{figure}
\section{Data}
\label{sec:data}
This analysis is based on the publicly available Millennium simulation
run  \citep[MS; ][]{Springel-05a,  Lemson-Springel-06}.  In  the first
part of  this section,  we review the  MS and the  semianalytic galaxy
modeling.  The  second part describes the  determination of background
densities and halo mass fractions.
\subsection{Millennium simulation}
\label{sec:mille}
The  MS  adopts  concordance  values  for the  parameters  of  a  flat
$\Lambda$  cold  dark  matter  cosmological model,  $\Omega_{\rm  dm}=
0.205$ and $\Omega_{\rm  b}= 0.045$ for the current  densities in cold
dark matter and baryons, $h= 0.73$ for the present dimensionless value
of  the Hubble  constant,  $\sigma_8=  0.9$ for  the  rms linear  mass
fluctuation in  a sphere of radius  $8 \hMpc$ extrapolated  to $z= 0$,
and $n= 1$ for the  slope of the primordial fluctuation spectrum.  The
simulation follows $2160^3$ dark  matter particles from $z=127$ to the
present day within a cubic  region $500\hMpc$ on a side. The resulting
individual particle mass  is $8.6\times10^8\hMsol$.  The gravitational
force has  a Plummer-equivalent  comoving softening of  $5\hkpc$.  The
Tree-PM $N$-body  code GADGET2  \citep{Springel-05b} has been  used to
carry out the simulation and the  full data are stored 64 times spaced
approximately equally in the logarithm of the expansion factor.

The halos are  found by a two-step procedure.  In  the first step, all
collapsed  halos with  at least  20 particles  are identified  using a
standard friends-of-friends (FoF)  group-finder with linking parameter
$b =  0.2$.  These  objects will be  referred to as  FoF-halos.  Then,
post-processing  with   the  substructure  algorithm   SUBFIND  \citep
{Springel-01} subdivides each  FoF halo into a set  of self-bound {\it
  sub-halos}.  Based on their assembly histories, individual sub-halos
are  populated  with  galaxies  by a  semi-analytic  prescription  and
various   observable  quantities  are   generated.   For   a  detailed
description of the semi-analytic galaxy catalog we refer the reader to
\cite{Croton-06}  and  \cite{DeLucia-Blaizot-07}.   The  stellar  mass
functions  in  this study  are  based  on the  DeLucia2006a\_SDSS2MASS
catalog         \citep[][{\tt         http://www.g-vo.org/Millennium}]
{Lemson-VirgoConsortium-06}.

The  following analysis  is based  on  FoF halos  which hereafter  are
addressed simply as halos.  In principle, one also could use different
halo definitions like those derived from spherical top-hat overdensity
criteria or  gravitational self-boundedness.  The former  would ease a
comparison with observations and the  latter would not suffer from the
bridging problem inherent to  the FoF approach.  However, according to
various  tests which  we  have  performed, the  results  based on  the
various  halo  identification  schemes  show little  difference.   The
advantage of  using FoF halos lies  in the simplicity  of the approach
which  makes  it a  very  common tool  for  the  analysis of  $N$-body
simulations.
\subsection{Background Density and Halo/Galaxy Mass Fraction}
\label{sec:density}
The MS database also provides  information on the global density field
besides  describing  individual halo  properties.   Here,  we use  the
densities, $\rho_{\rm CiC}$, which are based on a Counts in Cell (CiC)
approach   using   cubic   cells    of   $~\sim2\hMpc$   on   a   side
\citep[][]{Hockney-Eastwood-88}  and $\rho_{\rm G10}$  (hereafter {\it
  background density}), which are derived from the former by smoothing
them  with a  $10\hMpc$ Gaussian  kernel.  In  this study,  we utilize
$\rho_{\rm  CIC}$ to compute  the total  amount of  matter in  a given
volume  and calibrate  the  use of  the  $10\hMpc$ background  density
fields, $\rho_{\rm G10}$.  We  assume that $\rho_{\rm G10}$ represents
the linear density field.

To determine the total mass of halos per cell, we re-implement the CiC
approach only taking into account the particles belonging to FoF halos
above a given  mass limit.  If a halo crosses the  boundary of a cell,
only the mass  of the halo inside the cell  is counted.  This quantity
is   used   to   compute    the   halo   {\it   mass   fractions}   in
Section~\ref{sec:haback}.  For  the determination of the  halo and the
GMFs   (Section~\ref{sec:massfunc}   and  Section~\ref{s:stars}),   we
attribute the total mass of halos or galaxies to the cell within which
their centers are located.  Eventually,  each cell is assigned a total
mass in halos,  a total mass in stars (galaxies), a  total mass, and a
smoothed background density (which  after multiplication with the cell
volume corresponds to a mass as well).

The  upper left  and  the two  lower  panels of  Figure~\ref{fig:step}
display  a  slice  of   the  density  contrast,  $\delta_{\rm  G10}  =
(\rho_{\rm  G10}-\langle\rho\rangle) /  \langle\rho\rangle$  scaled by
the variance of the  matter fluctuations within a volume corresponding
to the Gaussian filter, $\sigma_{\rm G10}$, at redshifts $z=0$, $z=1$,
and  $z=3$.  Typical  patterns of  the large-scale  density  field are
apparent, such  as roughly spherical high  density regions, filaments,
and  voids.   According to  linear  theory,  $\delta/\sigma$ does  not
change with time. Indeed, the  scaled density contrast is very similar
in   shape    and   amplitude   at   the    redshifts   we   consider.
Figure~\ref{fig:dd}  shows  the  differential  mass  distributions  as
functions  of  the  scaled  density  contrast.  Red  curves  show  the
lognormal  fits  to  the  distribution  \citep[cf.,][]{Coles-Jones-91,
  Neyrinck-Szapudi-Szalay-09}.   The approximate invariability  of the
scaled contrast reaffirms the presumption of linearity of $\delta_{\rm
  G10}$.

The top right panel  of Figure~\ref{fig:step} shows the halo distribution
within a slice of $10\hMpc$ thickness at $z=0$ at the same location as
slice used  for the contrasts.  The  under-laid gray scale  image is a
replication  of the  contrast  on  the left.   As  expected, the  halo
distribution follows the pattern  of the background density field.  In
the  following, we  measure the  dependence  of the  fraction of  mass
captured in halos or galaxies as a function of the background density.
\begin{figure*}
\epsfig{file=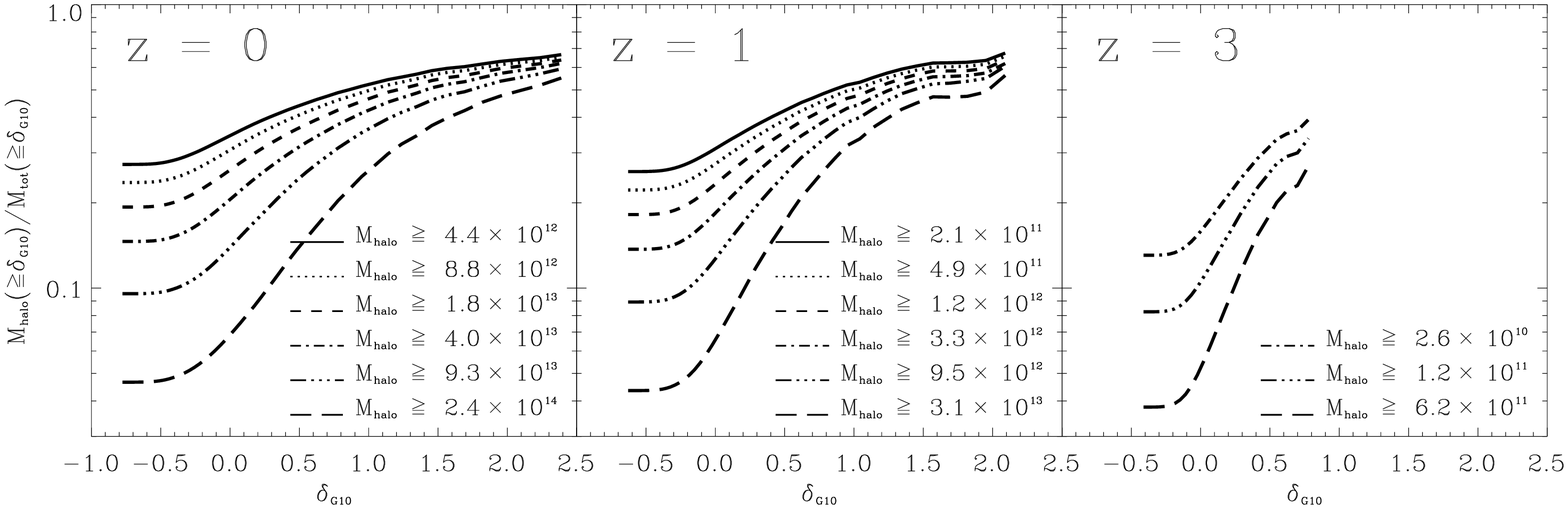, width = 0.95\hsize}
\caption{\label{fig:maiha} Halo  to total mass fraction  as a function
  of background density contrast  , $\delta_{\rm G10}$.  As indicated,
  the line  styles correspond to  different lower mass limits  for the
  halos used  to compute to total  halo mass budget. For  the left and
  middle panels, these masses correspond to the equivalent peak heights
  $\nu=  0.9$, $1.0$,  $1.2$, $1.4$,  $1.7$, and  $2.1$.  In  the $z=3$
  panel on the right only  the limiting masses corresponding to $\nu =
  1.4$, $1.7$,  and $2.1$  are shown.  At  that redshift,  $\nu$-values
  below $1.4$  correspond to halos  with less than  $\sim20$ particles
  which are not resolved.}
\end{figure*}
\begin{figure}
\epsfig{file=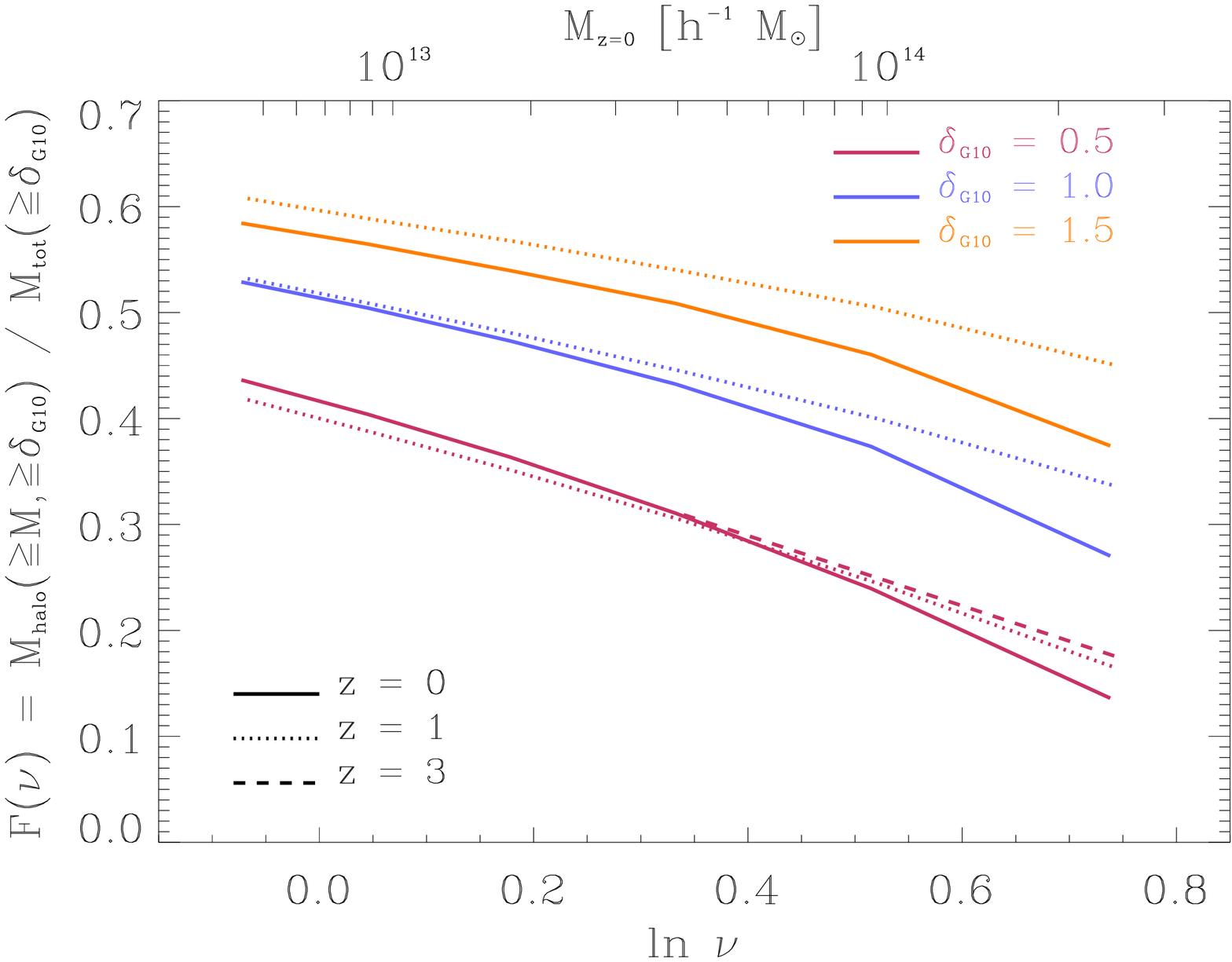, width = 0.95\hsize}
\caption{\label{fig:iso512}   Conditional    cumulative   halos   mass
  function,  $F(\nu)$, i.e.,  fraction of  mass locked  in  halos with
  respect to the total mass within a region confined by the contrasts,
  $\delta_{\rm G10}=0.5$, $1.0$, and $1.5$  as a function of halo mass
  given parameterized by the equivalent peak height $\nu = \delta_{\rm
    c}(z)/\sigma(M,z)$.   The  mass scale  at  the  top indicates  the
  equivalent  masses  for  $z=0$.   Solid, dotted,  and  dashed  lines
  correspond to redshifts $z=0$, $1$, and $3$, respectively.  At $z=3$
  $\delta_{\rm G10}$ does not reach $1$.}
\end{figure}
\section{Halo mass fractions}
\label{sec:massfrac}
The first part of this  section notes some general features associated
with  the characterization of  the density  field.  In  the subsequent
paragraph,  we examine  the dependence  of the  halo mass  fraction on
environment.   The {\it  halo mass  fraction} is  defined as  the mass
locked  in halos  divided by  the total  mass in  a given  volume.  As
stated  before, the  environment is  quantified based  on  the smoothed
density within this volume.  The halo mass fraction is closely related
to the  cumulative halo mass function  which will be  discussed in the
subsequent section.
\subsection{Mass within Isodensity Surfaces}
\label{sec:iso}
The  upper panel  in  Figure~\ref{fig:bg} shows  the  ratio between  the
average CiC-density of all cells located in regions of a given density
contrast, $\delta_{\rm  G10}$, as a  function of the  density contrast
itself. Line styles correspond  to different redshifts.  At a contrast
of $\delta_{\rm G10}\approx0.2$, the  mass within a cell approximately
corresponds to  the value of  $\rho_{\rm G10}$ multiplied by  the cell
volume.  For higher  contrasts, the actual mass deposited  in the cell
is  larger than  that deduced  from the  smoothed density  field.  The
opposite  is true  for contrasts  below $0.2$.   This  behavior simply
reflects that extremes are leveled by the smoothing procedure.

More  important for  the subsequent  analysis  is the  lower panel  in
Figure~\ref{fig:bg}.   It  displays   the  mean  value  of  $\rho_{\rm
  CIC}/\rho_{\rm  G10}$ within  a  volume confined  by the  isodensity
surface, $\rho_{\rm G10}$,  as a function of $\rho_{\rm  G10}$ or it's
equivalent $\delta_{\rm  G10}$.  The ratio,  $\rho_{\rm CIC}/\rho_{\rm
  G10}$, gives the average factor with which the confining density has
to be multiplied to get the  true mean density or the true mass within
the enclosed  volume. For different redshifts,  indicated by different
line  styles, one  finds slightly  different ratios,  but  the overall
behavior  is similar.  For  density contrasts  between $0$  and $1.5$,
which   are  of   interest  here,   one   obtains  $1\lesssim\rho_{\rm
  CIC}/\rho_{\rm G10}\lesssim3$.  This factor  has to be accounted for
when  the  total mass  within  a given  volume  is  inferred from  the
confining isodensity surface.  Loosely speaking,  a factor of 2 has to
be  multiplied  to  the   value  of  the  confining  surface  density,
$\rho_{\rm G10}$, to recover the mass inside.
\subsection{The dependence of the halo mass fraction on environment}
\label{sec:haback}
The  three panels in  Figure~\ref{fig:maiha} show  the {\it  halo mass
  fractions} for the redshifts $z=0$, $z=1$, and $z=3$.  The halo mass
fraction is defined  as the fraction of mass locked  in halos (above a
given limiting  mass) and the total  mass within a  volume confined by
isodensity surfaces  with a density  contrast $\delta_{\rm G10}$  as a
function of  that contrast. Line styles  refer to halo  mass limits as
indicated.  For  each redshift, these mass limits  correspond to fixed
values  of  the  equivalent   peak  height,  $\nu(M,z)  =  \delta_{\rm
  c}(z)/\sigma(M,z)$,   where   $\sigma(M,z)$   is  the   rms   linear
overdensity within a  sphere containing the mass $M$  at redshift $z$,
and  $\delta_{\rm  c}(z)$  is  the linear  overdensity  threshold  for
collapse at that  redshift.  The usage of $\nu$  instead of the actual
halo  mass should  remove  much  of the  cosmology  dependence of  our
results and  will be most useful  for the discussion of  the halo mass
functions  below. The  mass  limits  for the  left  and middle  panels
correspond to $\nu\approx 0.9$, $1.0$, $1.2$, $1.4$, $1.7$, and $2.1$.
In the  $z=3$ panel (on the  right) mass resolution only  allows us to
show the graphs for the  three highest $\nu$ values, $1.4$, $1.7$, and
$2.1$.

Figure~\ref{fig:iso512} shows the fraction of mass in halos with respect
to the total mass for confining density contrasts of $\delta_{\rm G10}
= 0.5$, $1.0$ and $1.5$ as a  function of the mass limit of the halos.
It is formally identical  to the conditional cumulative mass function,
$F(\nu)$, which is discussed in  the next section. However, here it is
computed based on the particle distributions of halos.  Therefore, the
mass  of halos which  transgress cell  boundaries is  accurately split
between the cells.  The analysis  in the next section is somewhat more
coarse in  the sense that  it attributes the  total mass of a  halo to
only  one cell,  namely  the cell  where  the center  of  the halo  is
located.   This can  introduce some  bias,  similar in  origin to  the
deviation  of   surface  and   mean  enclosed  density   discussed  in
~\ref{sec:iso}. However, the comparison  between the two methods yields
acceptable agreement  which implies  confidence in the  computation of
the mass functions presented below.
\subsection{Where are the Warm--Hot Baryons?}
Our  findings   have  consequences  for  the   detection  of  warm-hot
intergalactic baryons  (WHIM).  Cosmological simulations  predict that
some 50\%  of all the baryons locally  appear in the form  of gas with
temperatures between  $10^5$ and $10^7\rm  K$ \citep{Cen-Ostriker-99}.
The  main  process  responsible   for  heating  the  baryons  to  such
temperatures is  large scale structure formation and  thus the missing
component   is  thought   to  be   associated  with   high  densities.
Observational detection  of this component became subject  of a number
of studies, and  yet a full account of warm--hot  baryons has not been
reached.

In  a  subsequent study  \cite{Dave-01}  have  argued  that the  major
fraction   of  WHIM   is   located  outside   galaxy  groups.    Using
hydrodynamical simulations they have  shown that the total fraction of
WHIM associated with galaxy groups is between 10\% and 25\%, depending
on the  group mass cut  and that the  majority of the WHIM  resides at
mean densities  of $10<\delta<200$. In  this work, no attempt  has been
undertaken to  study the contribution  of baryons in galaxy  groups to
the WHIM  as a function of  environment. Here we aim  to determine the
fraction of WHIM located in groups at high background densities.

Figure~\ref{fig:maiha} suggests a dominant contribution of groups to the
matter budget at  high overdensities, resolving 60\% of  total mass in
groups with  mass exceeding $2\times10^{13}M_\odot$ and  70\% of total
mass in  groups with mass  exceeding $4\times10^{12}M_\odot$.  Without
any background  density restriction these groups  account for 20\%--30\%
of matter.

It is important to note that no gas component has been included in the
Millennium Simulation.  To derive conclusions  on the WHIM, we rely on
a model  describing the  distribution of gas  relative to  the overall
matter distribution which is equivalent to dark matter distribution in
the current context.  Here, we  assume for simplicity that gas follows
dark matter.  Hydrodynamical simulations  show that this assumption is
well  justified for  local  densities below  $\lesssim10^4$ times  the
cosmic  mean   density  \citep[cf.,[]{Faltenbacher-07}.   Furthermore,
  observational  and theoretical  accounts for  baryons  inside groups
  indicate  that the  universal  baryon fraction  is nearly  recovered
  \citep{Kravtsov-Nagai-Vikhlinin-05,Sun-09,Giodini-09,McGaugh-10},
  and potentially can be fully resolved by accounting the baryons near
  the virial  radius. Thus,  on the supposition  that gas  follows the
  dark matter  our findings  indicate that 20\%--30\%  of the  WHIM is
  located in  groups if no  background density constraint  is imposed,
  whereas 60\%--70\% of the WHIM  is residing in galaxy groups at high
  background  densities.   The former  agrees  well  with the  figures
  quoted in \cite{Dave-01}.

The only observational result published on the global fraction of mass
in groups, by \cite{Reiprich-Bohringer-02},  uses {\it ROSAT} All Sky
Survey data, and  quoted $\Omega_{cluster}=0.012$ (or 5\% contribution
of clusters  to the total  mass budget) for masses  exceeding $10^{14}
M_\odot h^{-1}$.   This measurement compares  well with our  value for
$M>2\times 10^{14} \hMsol$, which corresponds to the same $\nu$ when a
difference in  the $\sigma_8$ value between observed  universe and the
Millennium run is taken into account. This account will likely improve
soon   since   current  X-ray   surveys   can   access  masses   below
$10^{13}M_\odot$    \citep{Finoguenov-07,    Finoguenov-09}.     Also,
detections  of WHIM  in X-ray  emission  regions agree  well with  our
findings at high  background densities.  For example, \cite{Werner-08}
find that most  masses of baryons in the  A222/A223 complex are locked
within the halos, with WHIM contributing 10\%--20\%.

At  this point  we  would like  to  emphasize that  the definition  of
densities  adopted in  WHIM studies  is different  from that  one used
here.  Our results constrain the  WHIM aspect within the definition of
density      fields     typical      to      spectroscopic     surveys
\citep[e.g.,][]{Kovac-10}.  In  contrast the computation  of the local
density of the Lyman$_\alpha$ absorbers can only be derived from their
HI column  densities, resulting  in a non-trivial  role played  by the
absorber's size and shape.  In \cite{Penton-Stocke-Shull-04} the first
comparison to LSS  density has been provided, showing  the half of the
absorbers reside in voids.

Figure~\ref{fig:maiha} limits the importance  of the WHIM component in
regions of  high density (as defined on  a 10 Mpc scale)  which is not
associated with groups to 30\%.  Therefore, the situation is much more
favorable toward detecting missing  baryons in underdense regions, but
there the temperature of the gas will be quite low.  In fact the major
success  in resolving  the  missing baryons  has  been achieved  using
techniques looking for colder gas \citep{Penton-Stocke-Shull-04}.
\section{Cumulative halo mass functions}
\label{sec:massfunc}
Up to this  point, we have used the particle  distribution of the halos
to split their mass among the  cells they occupy.  In the following we
use the halo  position to attribute the entire halo  mass to one cell.
This reduces  the accuracy of  the determination of the  mass fraction
slightly ($\lesssim  5\%$).  However, that way the  treatment of halos
and  model galaxies  which  are examined  hereafter  and have  unknown
spatial extent can be matched.

Plenty of studies  have been devoted to investigate  the mass function
of  halos in  the cosmological  context.  Early  analytical approaches
suggested  that the  mass function  is universal,  i.e.\ its  shape is
independent of time and background cosmology if adequate variables are
used.   In   a  seminal  study,   \cite{Press-Schechter-74}  used  the
equivalent  peak  height,  $\nu  = \delta_{\rm  c}/\sigma$  \citep[see
  also,][]{Bond-91,  Lee-Shandarin-98, Sheth-Tormen-99}.   Such models
can   help  gain   insight  into   the  statistics   of  gravitational
collapse. However, these highly nonlinear processes are complex, and a
final  validation  of  the  models  can only  be  obtained  by  direct
comparison to  numerical simulations.  Based  on $N$-body simulations,
\cite{Jenkins-01} and \cite{Evrard-02}  presented fitting formulae for
the  differential mass  functions accurate  to  $\sim10\%-20\%$. These
studies supported  the view that mass functions  are indeed universal.
Also, the  results obtained by  \cite{Lukic-07} are consistent  with a
universal  mass  function;  however,   they  report  a  mild  redshift
dependence    at    low    redshifts   \citep[see    also][]{Reed-03}.
\cite{Warren-06} found  a fitting formula accurate to  $\sim5\%$ for a
fixed cosmology at  $z=0$. Recently, however, based on  a large set of
$N$-body   simulations    with   different   cosmological   parameters
\cite{Tinker-08}  revealed   that  the   mass  function  can   not  be
represented by a universal mass function at this level of accuracy. In
particular  they  found  that  the  amplitude  of  the  mass  function
decreases monotonically by $\approx  20\%-50\%$ from $z=0$ to $z=2.5$.
Several  studies   explore  the  mass  functions   at  high  redshifts
\citep[e.g.,][]{Reed-07, Cohn-White-08}  but they do  not particularly
focus on universality.

In  general, analytical and  numerical studies  preferentially discuss
the (unconditional) differential mass  function.  For our purposes the
integrated  form,  i.e.\   the  unconditional  cumulative  halos  mass
function (UHMF) is more useful since it directly gives the fraction of
mass in halos relative to the total mass. Additionally, the functional
forms of  the unconditional differential  mass functions, as  given in
the  literature, are  complex and  their integrals  are even  more so.
Thus, for  our purposes it seems  to be a better  strategy to directly
examine  the cumulative  mass  function and  derive  a simple  fitting
function for  it. Consequently, in the  first part of  this section we
introduce a fitting function for  the UHMFs.  In the second part, this
function  is  modified  to  be  applicable  for  the  cumulative  mass
functions at different background  densities, hereafter referred to as
conditional cumulative halo mass functions (CHMFs).
\begin{figure}
\epsfig{file=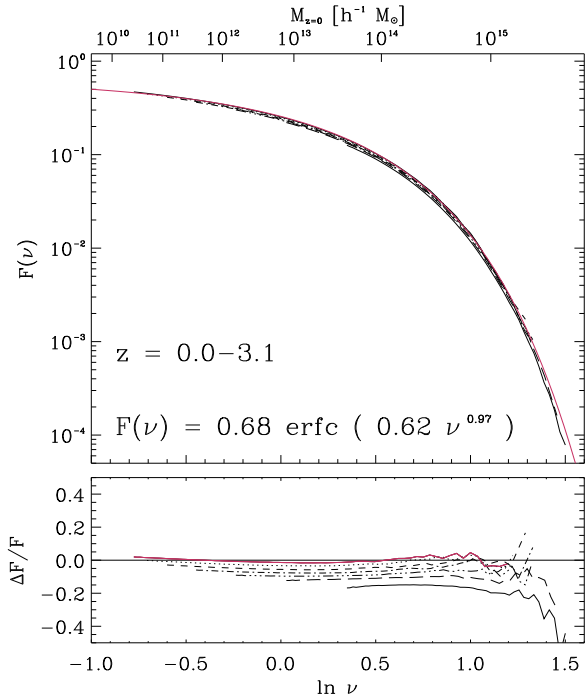, width = 0.95\hsize}
\caption{\label{fig:sf} Upper panel: cumulative  mass functions
  for redshifts $z=0.00$, $0.14$,  $0.36$, $0.69$, $1.17$, $1.91$, and
  $3.06$ shown as  solid, dotted, dashed, dot-dashed, three-dot-dashed,
  and  solid lines,  respectively. Halo  mass is  given parametrically
  through the  equivalent peak height  $\nu=\delta_{\rm c}/\sigma$. At
  the top  the corresponding mass  scale for $z=0$ is  indicated.  The
  red line presents a fit for the mass function at $z=0$.  The fitting
  function  including   the  parameters  is  quoted   in  the  bottom
  line. Lower panel: residuals of the mass functions and the fit
  at $z=0$. The  red line displays the residual  for the mass function
  at  $z=0$  which  has  been   used  for  fitting.   An  accuracy  of
  $\lesssim5\%$  is achieved  for  halo masses  between $10^{10}$  and
  $10^{15}\hMsol$.}
\end{figure}
\begin{figure*}
\epsfig{file=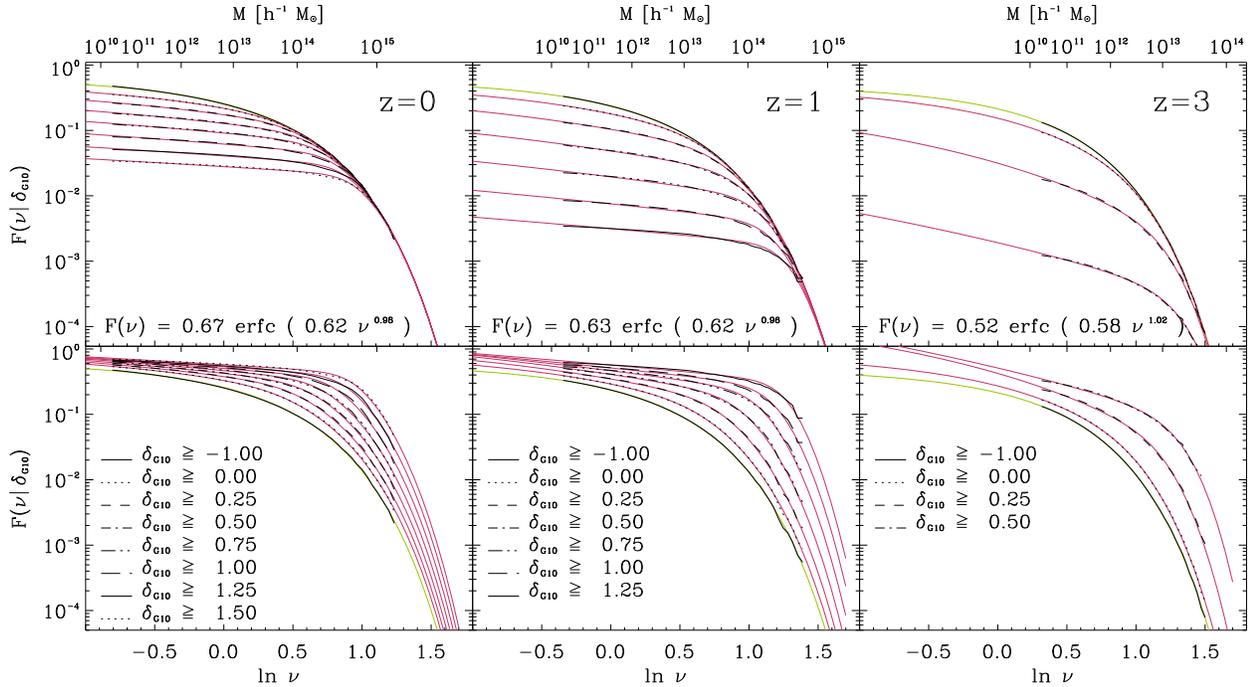, width = 0.95\hsize}
\caption{\label{fig:alion}  Dependence of  the halo  mass  fraction on
  environment  for the redshifts  $z=0$, $1$,  and $3$.   Lines styles
  correspond the  CHMFs in regions confined by  the density contrasts,
  $\delta_{\rm G10}$ as indicated. Green lines represent a fits to the
  UHMF using  Eq.~\ref{equ:fit}. The fitting  parameters are displayed
  at the bottom of the upper panels. Red lines show fits for the CHMFs
  using the fitting procedure described in Section~\ref{sec:cmf}.  The
  only  difference between  the upper  and  lower panels  is the  mass
  normalization.   In  the  upper  panels,  $F$ gives  the  halo  mass
  fraction with respect to the total  mass in the box and in the lower
  panels  with respect  to the  total mass  enclosed in  the overdense
  regions.}
\end{figure*}
\begin{figure}
\epsfig{file=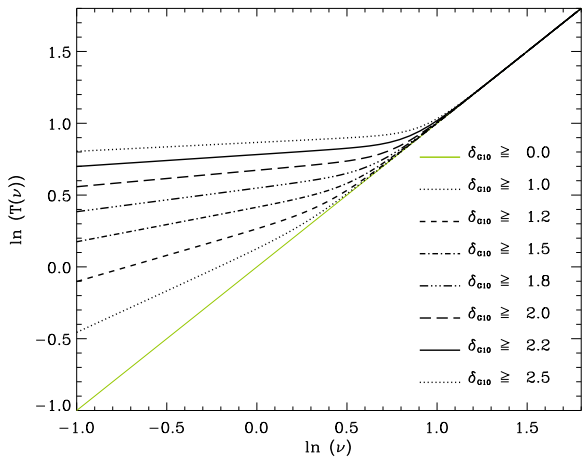, width = 0.95\hsize}
\caption{\label{fig:cotran} Coordinate  transformation used to convert
  unconditional  into  conditional mass  functions  for the  indicated
  density contrasts. To guide the eyes the green line for $\delta_{\rm
    G10}=0$  displays  the   identity.   For  higher  contrasts,  this
  transformation  keeps the  mass function  for large  values  of $\ln
  (\nu)$ unchanged yet it causes  the mass function at low $\ln (\nu)$
  to level off.}
\end{figure}
\begin{figure}
\epsfig{file=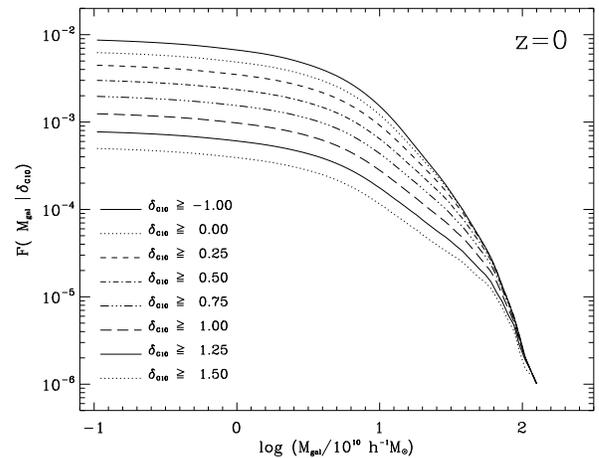, width = 0.95\hsize}
\caption{\label{fig:allstar}    Environmental   dependence    of   the
  cumulative  galaxy mass function.   Except for  the usage  of galaxy
  masses instead  of halo masses the  plot is equal to  the upper left
  panel in  Figure~\ref{fig:alion}; the  line styles are  adopted from
  there as well.}
\end{figure}
\begin{figure*}
\epsfig{file=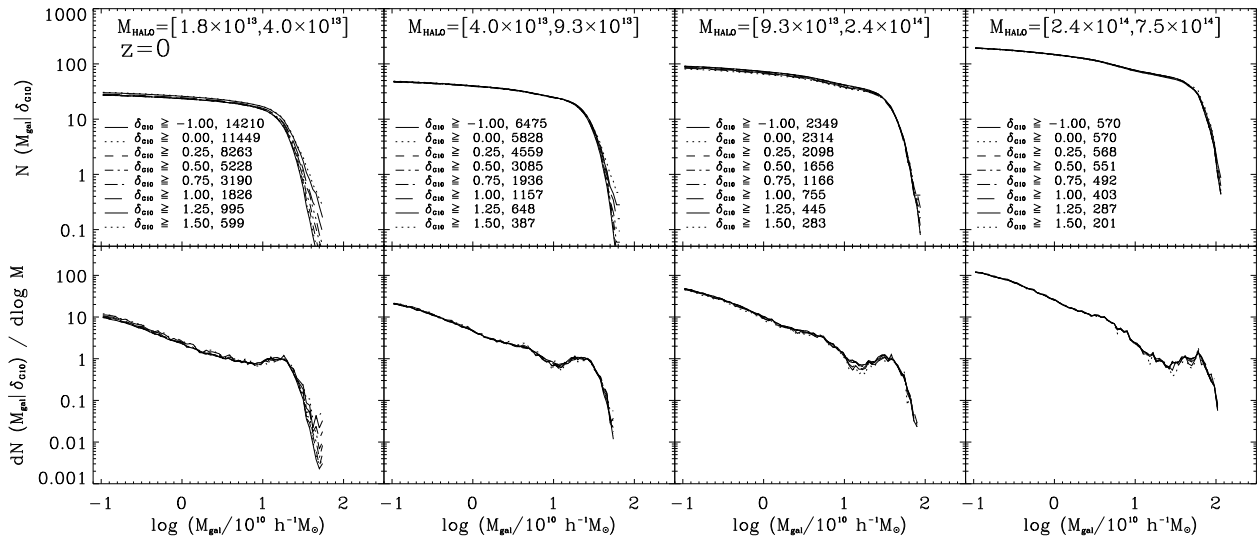, width = 0.95\hsize}
\caption{\label{fig:a8} Upper  panel: cumulative galaxy  mass function
  in   galaxy   groups  in   different   environments.  Lower   panel:
  differential  galaxy mass  function  in galaxy  groups in  different
  environments. On each panel, we denote both the density field and the
  number of embedded groups.}
\end{figure*}
\begin{figure*}
\epsfig{file=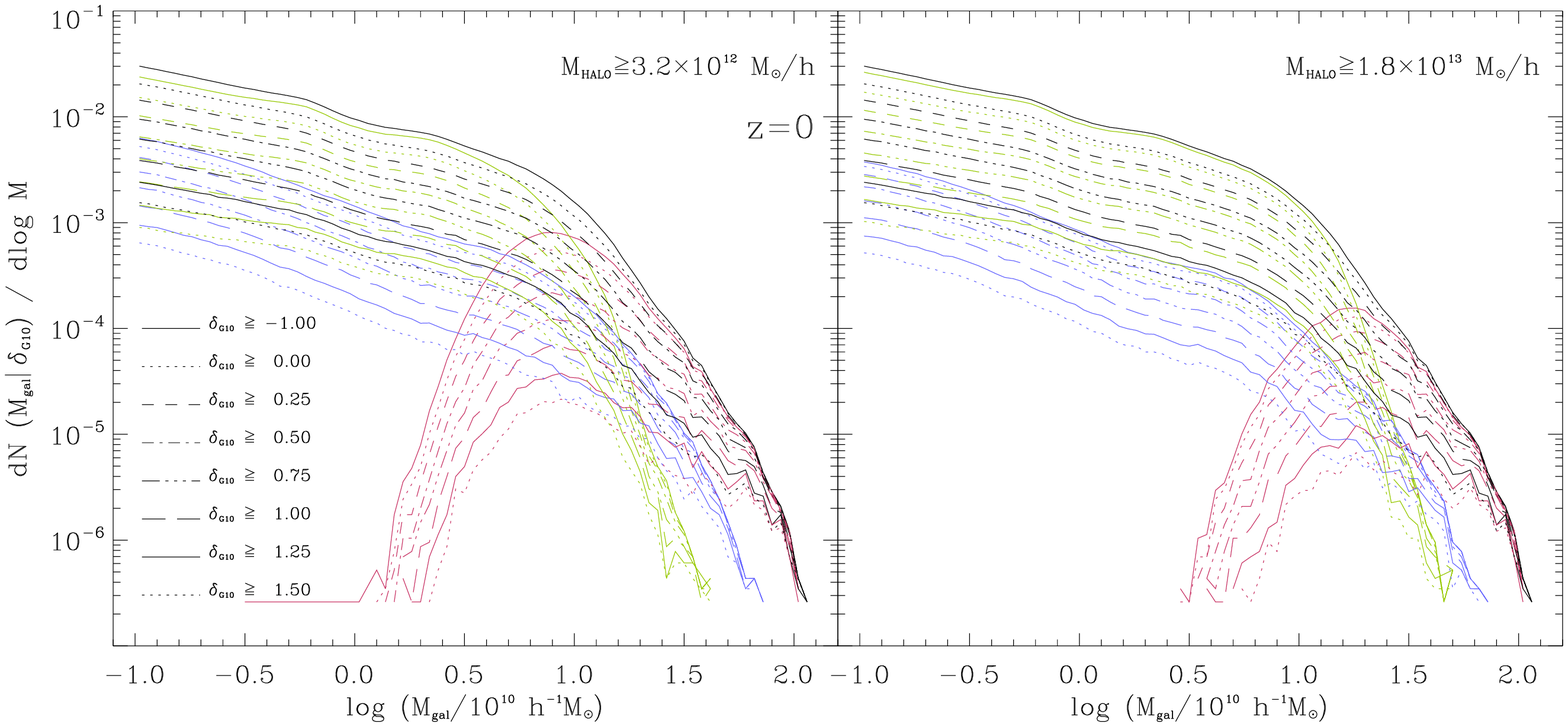, width = 0.95\hsize}
\caption{\label{fig:a2} Differential galaxy mass function in different
  environments  (black lines) separating  the contribution  of central
  (red lines)  and non-central galaxies (green lines)  in groups above
  the masses  indicated in the upper  right corner of  the panels. The
  green lines show the contribution of galaxies not belonging to those
  groups.   The  particular role  of  central  galaxies  is that  they
  provide a hump-like contribution to  the mass function and explain a
  variation of that hump with the environment.}
\end{figure*}
\subsection{Unconditional Cumulative Halo Mass Functions}
\label{sec:umf}
Under  the assumption  that the  initial density  field is  a Gaussian
random  field  and  the  validity  of the  spherical  collapse  model,
\cite{Press-Schechter-74}  derived the mass  fraction locked  in halos
above a given mass to the total mass in the universe,
\begin{equation}
\label{equ:PS}
F_{\rm PS}(\nu) = \erfc\left({\nu\over\sqrt{2}}\right),
\end{equation}
where   $\nu$  is  the   equivalent  peak   height  as   discussed  in
\$~\ref{sec:haback},  which, for  a given  cosmology, can  be uniquely
transformed  into  a  corresponding   halo  mass,  $M$.   $F(\nu)$  is
equivalent to the UHMF as introduced above.  \cite{Press-Schechter-74}
argued that  this parameterization makes the  mass function universal,
i.e.  independent of  evolutionary changes and cosmological parameters
which are covered by the time and cosmology dependence of $\nu$.  Over
the  last 20  years, high-resolution  N-body  simulations demonstrated
that Equation~\ref{equ:PS} reproduces the numerical halo mass function
qualitatively.  However, certain systematic deviations became apparent
\citep[e.g.,][]{Sheth-Tormen-02,  Warren-06, Tinker-08}.   Compared to
$N$-body results,  Equation~\ref{equ:PS} produces too  many halos with
masses  corresponding to  $\nu=1$ and  to few  at the  high  mass end.
Nevertheless, we  use the functional form  of Equation~\ref{equ:PS} as
template for our fitting formula:
\begin{equation}
\label{equ:fit}
F_{\rm fit}(\nu) = A\ \erfc\left({a\ \nu^{\ b}}\right),
\end{equation}
with  the fitting  parameters $A$,  $a$, and  $b$. The  upper  panel of
Figure~\ref{fig:sf} displays  UHMFs for  FoF halos derived from  the MS.
Different  line  styles  represent  UHMFs derived  from  snapshots  at
redshift $z=0.00$, $0.14$, $0.36$, $0.69$, $1.17$, $1.91$, and $3.06$.
The red  line shows  a fit  to the UHMF  at $z=0$.   The corresponding
fitting parameters are given in the panel.  For the computation of the
halo   masses,  we   applied   the  correction   formula  proposed   by
\cite{Warren-06}: $N{\rm corrected} = N(1-N^{0.6})$, where $N$ denotes
the number of  particles within a given FoF halo.   This correction is
most effective at  the low mass end. The fitting  range is confined by
FoF halos with  masses between $2\times10^{10}\hMsol$  (200 particles)
and $10^{15}\hMsol$.

The lower panel of Figure~\ref{fig:sf} shows the residuals between the
UHMFs  at the  given redshifts  and the  fit based  on the  $z=0$ mass
function. The red line highlights the residual between the fit and the
$z=0$  mass function. At  $z=0$ and  for masses  in the  range between
$10^{10}\hMsol$ and  $10^{15}\hMsol$, the fit  is accurate to  the 5\%
level.   However, we  note an  increasing offset  with  redshift which
results in  a deviation  of 15\% at  $z=3.06$.  Similar  findings have
been     reported     in     \cite{Tinker-08}.     The     expression,
Equation~\ref{fig:sf},  is  an  excellent  fitting  function  for  the
cumulative  mass   function  at   $z=0$.  It  is   ``universal'',  i.e.,
independent of cosmology at a  degree to which the parameterization by
the equivalent peak height, $\nu$, permits.
\subsection{Conditional Cumulative Halo Mass Function}
\label{sec:cmf}
The term  ``conditional mass function''  has been used to  address two
related problems: (1) to  describe the mass distribution of progenitor
halos which end up in a given halo at later times; (2) to refer to the
mass  distribution of  halos within  a  region of  a given  background
density    \citep[cf.,][]{Mo-White-96,   Sheth-98,   Sheth-Lemson-99b,
  Sheth-Tormen-02} .  Here we  will investigate the latter.  According
to excursion set theory, the  basic equation to approach such problems
is
\begin{equation}
\label{eq:excurs}
F(M|\delta_0,\sigma_0)   =   \erfc\left({1\over\sqrt{2}}  {\delta_{\rm
    c}-\delta_0\over\sigma-\sigma_0}\right),
\end{equation}
where  $\delta_{\rm  c}$  and  $\sigma$   are  the  same  as  used  in
Equation~\ref{equ:PS} and  $\delta_0$ and $\sigma_0$ are  the linear density
contrast and the dispersion of the background density field.  It gives
the  fraction of  mass in  collapsed halos  of mass  greater  than $M$
(corresponding to $\nu=\delta_{\rm c}/\sigma$)  in a region that has a
linear density contrast  $\delta_0$ \citep[for a comprehensive review
  including  references see,][]{Zentner-07}.   In the  current context,
two difficulties  arise when this  equation is to be  applied.  First,
the boundary surface and hence the volume of the overdense regions can
be quite  irregular which makes  it difficult to  determine $\sigma_0$
appropriately.    Second,   to    facilitate   the   comparison   with
observations, we  intend to compute  the mass fraction  within regions
{\it above} a given $\delta_{\rm lim}$, in other words cumulative with
respect to the  background density.  However Equation~\ref{eq:excurs} gives
the  halo  mass fraction  {\it  at}  a  given $\delta_0$.   Therefore,
obtaining  an expression suitable  for our  purposes would  require an
integration  of Equation~\ref{eq:excurs} with  respect to  $\delta_0$. This
integration leads to a lengthy  expression, which is too complex to be
used  as a  model  for a  fitting formula  (as  done for  the UHMF  in
Section~\ref{sec:umf}).
 
Therefore, we  choose a  more phenomenological approach  starting with
the inspection  of Figure~\ref{fig:alion}.  Black  lines display UHMFs
and CHMFs  in regions above a  given density contrast  at redshifts of
$z=0$, $z=1$, and  $z=3$.  The CHMFs are determined  by summing up the
mass  of  halos   which  reside  in  cells  above   a  given  contrast
$\delta_{G10}$.  The difference between  the upper and lower panels is
only in normalization.  In the upper panels, we use  the total mass in
the box whereas  in the lower panels we use the  total mass within the
cells above $\delta_{G10}$.  Therefore,  the CHMFs in the upper panels
lie systematically below the UHMFs;  this is because a fraction of the
total volume  is excluded by the  density criterion -- and  so are the
halos in it -- but the mass fraction is still computed with respect to
the  total mass  in  the box.   In  the lower  panels,  the CHMFs  lie
systematically above  the UHMFs  which is a  result of  the decreasing
total  mass in  the volume  above the  given background  density.  The
latter is  more physical but the  former illustrates the  fact that at
the high-mass  end all  CHMFs display the  same behavior as  the UHMF.
The  green lines  show fits  to the  UHMF. The  values of  the fitting
parameters are given at the bottom of the upper panels (here the UHMFs
are fitted separately  for each redshift).  The red  lines are fits to
the CHMFs.

\begin{table}
\begin{center}
\begin{tabular}{|c|c|c|c|c|c|}
  \hline   Redshift&$\delta_{\rm  G10}$&$\Delta$&$a$&$b$&$f_M$\\\hline
  $z=0$& 0.00 &  1.00 & 1.00 & 1.00 &  1.00 \\ & 1.00 &  0.58 & 7.76 &
  0.28 & 1.51 \\ & 1.25 & 0.36 & 7.90 & 0.41 & 2.18 \\ & 1.50 & 0.24 &
  8.45 & 0.55 & 3.30 \\ & 1.75 & 0.16 & 8.91 & 0.66 & 5.12 \\ & 2.00 &
  0.11 & 9.77  & 0.76 & 8.16 \\ &  2.25 & 0.08 & 10.25  & 0.85 & 13.17
  \\ & 2.50 & 0.06 & 10.87  & 0.92 & 20.55 \\\hline $z=1$& 0.00 & 1.00
  & 1.00 & 1.00 & 1.00 \\ & 1.00 & 0.61 & 7.58 & 0.51 & 1.63 \\ & 1.25
  & 0.34 &  7.28 & 0.71 & 3.34 \\  & 1.50 & 0.20 & 7.79  & 0.90 & 8.41
  \\ & 1.75  & 0.12 & 8.69  & 1.05 & 23.86 \\  & 2.00 & 0.07  & 8.72 &
  1.17 & 67.96 \\ & 2.25 & 0.05 & 10.18 & 1.27 &180.57 \\\hline $z=3$&
  0.00 & 1.00  & 1.00 & 1.00 & 1.00  \\ & 1.00 & 0.74 &  7.37 & 0.91 &
  1.77 \\ & 1.25 & 0.32 & 7.27 &  1.20 & 13.43 \\ & 1.50 & 0.12 & 6.18
  & 1.41 &266.83 \\\hline
\end{tabular}
\caption{\label{tab:cmf} CHMF fitting parameters, $\Delta$,$a$,$b$ and
  mass  factor  $f_{m}$  for  the background  densities,  $\delta_{\rm
    G10}$, and redshifts $z=0$, $1$, and $3$.}
\end{center}
\end{table}
Guided   by    the   behavior   seen   in   the    upper   panels   of
Figure~\ref{fig:alion},  we conceive  a  fitting formula  for the  CHMFs
which  is based  on a  parameter-dependent  coordinate transformation,
$T(\nu)$,  such that  for an  adequate set  of  parameters $F(T(\nu))$
(here  $F$  denotes  the UHMF)  matches  a  given  CHMF.  We  use  the
following  functional form for  these transformations,  constructed to
map high $\nu$ values onto  themselves but narrowing the range for low
$\nu$s:
\begin{equation}
  \label{equ:tran}
  \ln (T(\nu)) = \Delta + {a x + (1.0-a)\ \ln[1.0+\exp(x)]\over b},
\end{equation}
where $x = b\ (\ln(\nu)-\Delta)$.  The values for the three parameters
$\Delta$,  $a$, and  $b$ are  listed in  Table~\ref{tab:cmf}  with the
corresponding  fits  shown  as  red lines  in  Figure~\ref{fig:alion}.
Figure~\ref{fig:cotran} illustrates  the coordinate transformations at
$z=0$.   $\Delta$  determines  where  the  deviation  from  one-to-one
correspondence takes place, $a$ gives the left hand side slope and $b$
defines the  smoothness of the  transition between the two  slopes. So
far, we  have described how  to fit the  CHMFs in the upper  panels of
Figure~\ref{fig:alion}.  To get the  mass normalization right, i.e., to
transform the  fits of the  upper panels into  the those of  the lower
panels, these  $F(T(\nu))$ have  to be multiplied  by the  mass factor
$f_M$ which is listed  in the rightmost column of Table~\ref{tab:cmf}.
It gives the ratio  of the total mass in the box  to the mass confined
to the considered overdense regions.

Our  findings may  help  to  illustrate some  of  the main  mechanisms
shaping the  halo mass functions  in different environments.   The top
heavy shape  of the CHMFs is caused  by a relative lack  of small mass
halos at high background densities.   For our way of parameterizing the
background density,  namely, by using isodensity  surfaces to indicate
all  the volume  interior  to it,  the  shape of  the cumulative  mass
functions at the high mass end is independent of environment. The only
difference  is induced by  the normalization  which reflects  the mass
confined to the high density  regions.  It causes the amplitude of the
mass function to rise.

As a concluding  remark we note that the use  of cumulative instead of
differential mass functions proved  valuable to extract these results.
First, because it directly gives the fraction of mass in halos above a
given mass with respect to the total mass.  In addition, it's shape is
simple compared to the  differential mass function which eases finding
a suitable fitting  strategy.  Finally, it allows us  to easily derive
some of  its basic  properties by analytically.   For instance,  it is
obvious that the ``wrongly normalized'' CHMFs in the upper panels must
coincide at the  high mass end.  Similarly easy to  derive is that the
fits  in  the   lower  panels  must  be  confined   by  $1$  even  for
extrapolations  to  smallest  halo  masses.   The fact  that  this  is
obviously  not the case  for two  of the  $z=3$ fits  demonstrates the
limits of the fitting procedure.   However, in general the fits behave
well, i.e., the  amplitudes remain $<1$ even if  extrapolated to small
masses.
\section{Conditional galaxy mass function}
\label{s:stars}
The semi-analytical  modeling of galaxies included in  the MS database
\citep{DeLucia-Blaizot-07,Lemson-VirgoConsortium-06}  provides stellar
masses for each model galaxy.   This allows us to compute the conditional
cumulative galaxy stellar mass function (CGMF) in exactly the same way
as the CHMFs have been determined, namely by summing up the mass of all
galaxies which reside in  cells above a given contrast $\delta_{G10}$.
We  set  the  lower  limit  for the  galaxy  masses  to  $10^9\hMsol$.
Figure~\ref{fig:allstar} shows the results of the CGMF normalized by the
total mass  in the  box, which  is the reason  why all  mass functions
coincide at  the high-mass  end, equivalently to  the behavior  of the
CHMFs.  In general, the changes in CGMF are qualitatively very similar
to that of the CHMFs: becoming top heavy at high background densities.

The qualitative similarities between  the CHMFs and CGMFs suggest that
the dependence  of the  model galaxy mass  function on  environment is
caused by  the corresponding dependence  of the halo mass  function on
environment,  rather than on  direct impact  of environment  on galaxy
evolution (generally  referred to as  nurture effects).  To  test this
conjecture, Figure~\ref{fig:a8} shows  the model galaxy mass functions
at different background densities with the additional restriction that
they reside in dark matter halos of a given mass range as indicated by
the labels right on top.   The upper panels show the cumulative galaxy
mass functions  and the  lower panels depict  the GMF  in differential
form.   The line  styles  correspond to  different density  thresholds
listed  in the  figure.  The  integer numbers  indicate the  number of
halos  in  those regions.   Evidently,  these  numbers  show a  strong
dependence  on environment  assuring  that the  galaxy mass  functions
displayed in a  single panel are based on very  different sets of host
halos.   Nevertheless, the  resulting mass  functions deviate  by less
than  $10$\% for galaxy  masses $\lesssim  10^{11}\hMsol$.  Therefore,
the large changes in the CGMFs can almost entirely be accounted for by
changes in the host halo population.  Nurture effects may have a minor
impact  on  shaping  the  model  galaxy  mass  function  in  different
environments.

There  are some  shortcomings in  the semianalytical  model  of galaxy
formation used here \citep{DeLucia-Blaizot-07}. The two most important
are: (1) cooling is instantaneously  shut down for galaxies whose halo
enters a  larger one; (2)  tidal forces are  not allowed to  strip off
stars reducing the luminosity  of a given galaxy.  Nevertheless, these
processes do mostly affect  satellite galaxies.  Thus, we believe that
the  behavior seen  at the  high mass  end of  the CGMFs  should  be a
reliable prediction from the semianalytical model.

To provide a further  illustration, we display the differential galaxy
mass function as a function of environment separating the contribution
of central  and non-central  galaxies in groups  in Figure~\ref{fig:a2}.
The  particular  role of  central  galaxies  is  that they  provide  a
hump-like  contribution  to  the  mass  function  and  illustrate  the
variation of that hump with the environment, therefore explain another
important    observational    result   \citep[e.g.,][]{Bolzonella-09}.
Explaining the hump-like  feature as being due to  the contribution of
central galaxies (and hence a bimodal GMF with centrals and satellites
as its  two constituents) is  somewhat different than  the explanation
given by  \citet{Bolzonella-09}, namely  that the hump  is due  to the
contributions of red galaxies to the  total GMF. In our view, the hump
is a consequence  of halo assembly, not of  the transformation of blue
into red galaxies,  although the two processes might  be linked in the
sense  that  the assembly  of  central  galaxies  might also  lead  to
quenching of their star formation  in some halos. We therefore predict
that  the hump  will also  be seen  in the  blue galaxy  mass function
alone, not only  in the total GMF, since many  halos host giant (blue)
spiral galaxies  as their central  galaxy. Indeed, this  bimodality in
the blue galaxy mass function has been observed by \cite{Drory-09}.

The thermodynamical  state of  the accreting gas  is expected  to vary
with redshift, enabling a cold  accretion at $z>2$ and subsequent star
formation in the  central galaxy even in halos  as massive as $10^{13}
M_\odot$ \citep{Dekel-09}. These are  expected to be the highest peaks
of density field at those redshifts and therefore obey our predictions
for the  behavior in high density  environments.  Observational search
for the transitional halo mass  between the cold and the hot accretion
mode is difficult, as this mass  is well below the sensitivity of both
spectroscopic   and   X-ray    surveys   for   defining   the   galaxy
groups. Instead, here  we propose to use the shape  of the galaxy mass
function to determine at which halo mass the transition occurs.  Since
the central  galaxies of groups  make a hump-like contribution  to the
galaxy mass  function, the blue  central galaxies should  constitute a
blue hump.  The location of the  blue hump in the galaxy mass function
can therefore be used through a comparison to numerical simulations to
determine the  transition halo mass  scale for shutting down  the star
formation or environmental dependence of cold accretion mode in galaxy
formation.   In contrast  to clustering  studies, the  proposed method
does not  induce a requirement on  the data to  be representative, and
can therefore  be applied to  a field of  any size or even  a selected
object, like a high-redshift supercluster.
\section{Conclusions}
\label{sec:conclusion}
Using the Millennium Simulation  and current schemes for density field
reconstruction,  we parameterize the  conditional halo  mass function.
As an  application, we  consider the role  of halos in  explaining the
missing baryon problem and the environmental dependence of galaxy mass
functions.  We  show that  in high-density environments  galaxy groups
provide a major contribution to  total matter content.  We discuss the
implication of this  result for search of missing  baryons.  Under the
well-justified  assumption that  baryons follow  dark matter,  we show
that its  amount can be  constrained using the observations  of galaxy
groups.  We  also point out  that the environmental changes  in galaxy
mass functions are  caused by changes in mass  function of groups and,
in particular,  that a hump-like  features in galaxy mass  function is
produced by the central galaxies of groups.
\section*{Acknowledgments} 
The  authors are  thankful  to the  anonymous  referee for  insightful
suggestions, to Simon White for  valuable comments on the paper and to
Eyal  Neistein  for  helpful  discussions.   Andreas  Faltenbacher  is
supported  by the  SA SKA  bursay  program and  acknowledges the  kind
hospitality  at   South  African  Astronomical   Observatory.   Alexis
Finoguenov acknowledges support from  {\it Spitzer} UDS Legacy program
to UMBC.  The  Millennium Simulation databases used in  this paper and
the web  application providing online access to  them were constructed
as  part  of  the  activities  of  the  German  Astrophysical  Virtual
Observatory.  The  authors thank  Gerard Lemson for  his help  with MS
database and the comments on the manuscript.
%

\end{document}